\documentclass[twocolumn,showpacs,preprintnumbers,amsmath,amssymb,superscriptaddress]{revtex4}

\usepackage{graphicx}
\usepackage{dcolumn}
\usepackage{bm}

\begin{document}

{Detailed version of SCIENCE \textbf{323}, 919 (2009).}

{http://dx.doi.org/10.1126/science.1167733}


\title{Observation of topologically protected Dirac spin-textures and $\pi$ Berry's phase in pure Antimony (Sb) and topological insulator BiSb}


\author{D. Hsieh}
\affiliation{Joseph Henry Laboratories of Physics, Princeton
University, Princeton, NJ 08544, USA}
\author{Y. Xia}
\affiliation{Joseph Henry Laboratories of Physics, Princeton
University, Princeton, NJ 08544, USA}
\author{L. Wray}
\affiliation{Joseph Henry Laboratories of Physics, Princeton
University, Princeton, NJ 08544, USA}
\author{D. Qian}
\affiliation{Joseph Henry Laboratories of Physics, Princeton
University, Princeton, NJ 08544, USA}
\author{A. Pal}
\affiliation{Joseph Henry Laboratories of Physics, Princeton
University, Princeton, NJ 08544, USA}
\author{J. H. Dil}
\affiliation{Swiss Light Source, Paul Scherrer Institute, CH-5232,
Villigen, Switzerland} \affiliation{Physik-Institut, Universit\"{a}t
Z\"{u}rich-Irchel, 8057 Z\"{u}rich, Switzerland}
\author{F. Meier}
\affiliation{Swiss Light Source, Paul Scherrer Institute, CH-5232,
Villigen, Switzerland} \affiliation{Physik-Institut, Universit\"{a}t
Z\"{u}rich-Irchel, 8057 Z\"{u}rich, Switzerland}
\author{J. Osterwalder}
\affiliation{Physik-Institut, Universit\"{a}t Z\"{u}rich-Irchel,
8057 Z\"{u}rich, Switzerland}
\author{G. Bihlmayer}
\affiliation{Institut f\"{u}r Festk\"{o}rperforschung,
Forschungszentrum J\"{u}lich, D-52425 J\"{u}lich, Germany}
\author{C. L. Kane}
\affiliation{Department of Physics and Astronomy, University of
Pennsylvania, Philadelphia, PA 19104, USA}
\author{Y. S. Hor}
\affiliation{Department of Chemistry, Princeton University,
Princeton, NJ 08544, USA}
\author{R. J. Cava}
\affiliation{Department of Chemistry, Princeton University,
Princeton, NJ 08544, USA}
\author{M. Z. Hasan}
\affiliation{Joseph Henry Laboratories of Physics, Princeton
University, Princeton, NJ 08544, USA} \affiliation{Princeton Center
for Complex Materials, Princeton University, Princeton, NJ 08544,
USA}\email{mzhasan@Princeton.edu}

\date{\today}

\begin{abstract}
A topologically ordered material is characterized by a rare quantum
organization of electrons that evades the conventional spontaneously
broken symmetry based classification of condensed matter. Exotic
spin transport phenomena such as the dissipationless quantum spin
Hall effect have been speculated to originate from a novel
topological order whose identification requires a spin sensitive
measurement. Using \textit{spin-resolved ARPES}, we probe the spin degrees of freedom and
demonstrate that topological quantum numbers are uniquely determined
from spin texture imaging measurements. Applying this method to pure antimony (Sb)
and Bi$_{1-x}$Sb$_x$, we identify the origin of its novel order and unusual chiral topological properties. These results taken together constitute the observation of surface electrons collectively carrying a geometrical quantum (Berry's) phase and definite chirality in pure Antimony, Sb, and topological insulator BiSb, which are the key electronic properties for realizing topological quantum computing via the Majorana fermion framework. This paper contains the details of our previously reported (first reported in Science \textbf{323}, 919 (2009)) observation of a negative mirror Chern quantum number for pure Sb.
\end{abstract}

\maketitle

Ordered phases of matter such as a superfluid or a ferromagnet are
usually associated with the breaking of a symmetry and are
characterized by a local order parameter [1], and the typical
experimental probes of these systems are sensitive to order
parameters. The discovery of the quantum Hall effects in the 1980s
revealed a new and rare type of order that is derived from an
organized collective quantum motion of electrons [2-4]. These
so-called ``topologically ordered phases" do not exhibit any
symmetry breaking and are characterized by a topological number [5]
as opposed to a local order parameter. The classic experimental
probe of topological quantum numbers is magneto-transport, where
measurements of the quantization of Hall conductivity $\sigma_{xy} =
ne^{2}/h$ (where $e$ is the electric charge and $h$ is Planck's
constant) reveals the value of the topological number $n$ that
characterizes the quantum Hall effect state [6].

\begin{figure*}
\includegraphics[scale=0.58,clip=true, viewport=0.0in 0.5in 11.3in 6.5in]{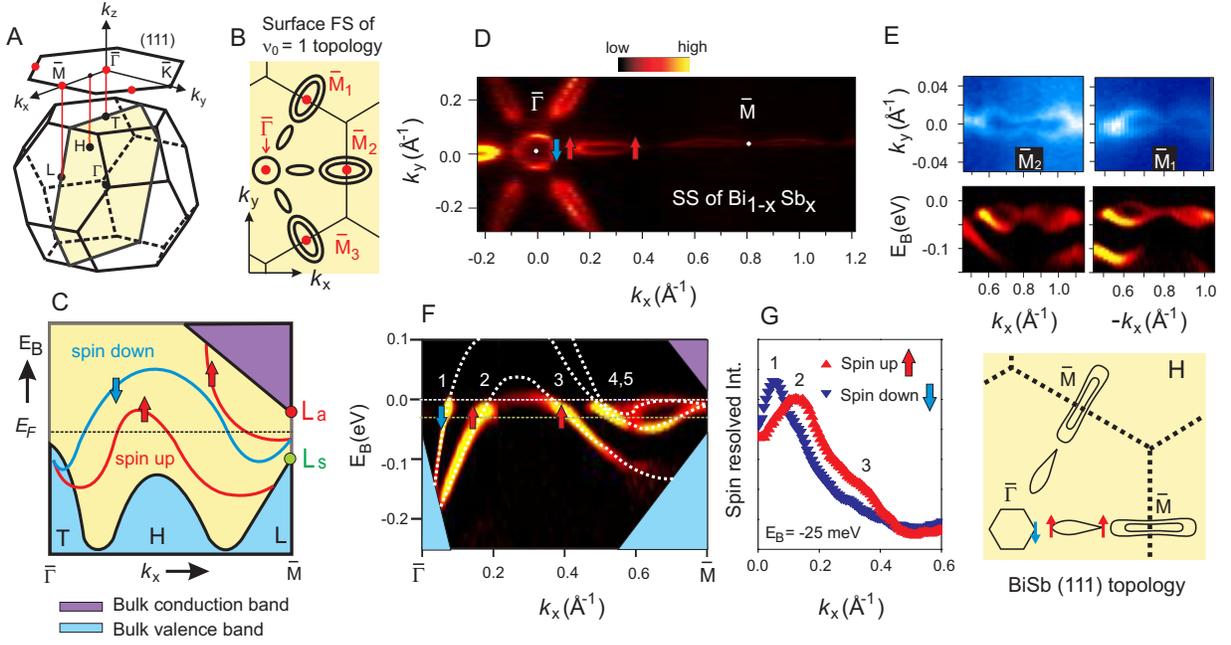}
\caption{Spin-texture and $\pi$ Berry's Phase in BiSb}
\end{figure*}

Recent theoretical and experimental studies suggest that a new class
of quantum Hall-like topological phases can exist in spin-orbit
materials without external magnetic fields, with interest centering
on two examples, the ``quantum spin Hall insulator" [7-9] and the
``strong topological insulator" [10,11]. Their topological order is
believed to give rise to unconventional spin physics at the sample
edges or surfaces with potential applications ranging from
dissipationless spin currents [12] to topological (fault-tolerant)
quantum computing [13]. However, unlike conventional quantum Hall
systems, these novel topological phases do not necessarily exhibit a
quantized charge or spin response ($\sigma_{xy} \neq ne^{2}/h$)
[14,15]. In fact, the spin polarization is not a conserved quantity
in a spin-orbit material. Thus, their topological quantum numbers,
the analogues of $n$, cannot be measured via the classic von
Klitzing-type [2] transport methods.

Here we show that spin-resolved angle-resolved photoemission
spectroscopy (spin-ARPES) can perform analogous measurements for
topological metals and insulators. We measured all of the
topological numbers for Bi$_{1-x}$Sb$_x$ and provide an
identification of its spin-texture, which heretofore was unmeasured
despite its surface states having been observed [10]. The measured
spin texture reveals the existence of a non-zero geometrical quantum
phase (Berry's phase [16,17]) and the handedness or chiral
properties. More importantly, this technique enables us to
investigate aspects of the metallic regime of the Bi$_{1-x}$Sb$_x$
series, such as spin properties in pure Sb, which are necessary to
determine the microscopic origin of topological order. Our
measurements on pure metallic Sb show that its surface carries a
geometrical (Berry's) phase and chirality property unlike the
conventional spin-orbit metals such as gold (Au), which has zero net
Berry's phase and no net chirality [18].

Strong topological materials are distinguished from ordinary
materials such as gold by a topological quantum number, $\nu_0$ = 1
or 0 respectively [14,15]. For Bi$_{1-x}$Sb$_x$, theory has shown
that $\nu_0$ is determined solely by the character of the bulk
electronic wave functions at the $L$ point in the three-dimensional
(3D) Brillouin zone (BZ). When the lowest energy conduction band
state is composed of an antisymmetric combination of atomic $p$-type
orbitals ($L_a$) and the highest energy valence band state is
composed of a symmetric combination ($L_s$), then $\nu_0$ = 1, and
vice versa for $\nu_0$ = 0 [11]. Although the bonding nature
(parity) of the states at $L$ is not revealed in a measurement of
the bulk band structure, the value of $\nu_0$ can be determined from
the spin-textures of the surface bands that form when the bulk is
terminated. In particular, a $\nu_0$ = 1 topology requires the
terminated surface to have a Fermi surface (FS) [1] that supports a
non-zero Berry's phase (odd as opposed to even multiple of $\pi$),
which is not realizable in an ordinary spin-orbit material.

\begin{figure*}
\includegraphics[scale=0.6,clip=true, viewport=0.0in 0.7in 11.0in 5.4in]{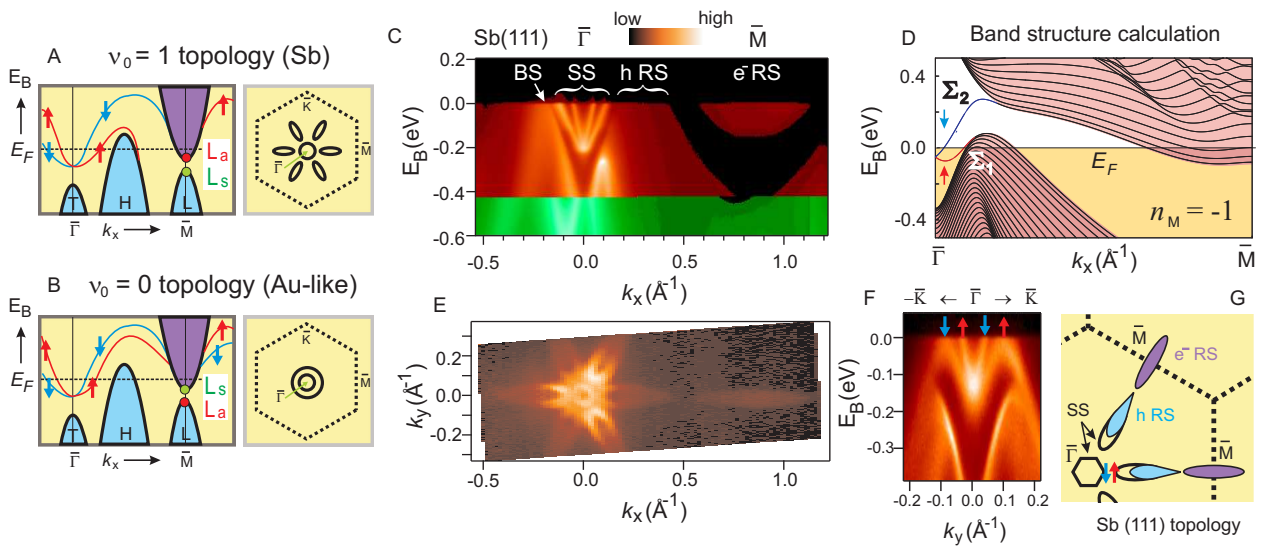}
\caption{Topological nature of the surface states in pure Sb (Antimony)}
\end{figure*}

In a general inversion symmetric spin-orbit insulator, the bulk
states are spin degenerate because of a combination of space
inversion symmetry $[E(\vec{k},\uparrow) = E(-\vec{k},\uparrow)]$
and time reversal symmetry $[E(\vec{k},\uparrow) =
E(-\vec{k},\downarrow)]$. Because space inversion symmetry is broken
at the terminated surface, the spin degeneracy of surface bands can
be lifted by the spin-orbit interaction [19-21]. However, according
to Kramers theorem [16], they must remain spin degenerate at four
special time reversal invariant momenta ($\vec{k}_T$ =
$\bar{\Gamma}$, \={M}) in the surface BZ [11], which for the (111)
surface of Bi$_{1-x}$Sb$_x$ are located at $\bar{\Gamma}$ and three
equivalent \={M} points [see Fig.1(A)].

Depending on whether $\nu_0$ equals 0 or 1, the Fermi surface
pockets formed by the surface bands will enclose the four
$\vec{k}_T$ an even or odd number of times respectively. If a Fermi
surface pocket does not enclose $\vec{k}_T$ (= $\bar{\Gamma}$,
\={M}), it is irrelevant for the topology [11,20]. Because the wave
function of a single electron spin acquires a geometric phase factor
of $\pi$ [16] as it evolves by 360$^{\circ}$ in momentum space along
a Fermi contour enclosing a $\vec{k}_T$, an odd number of Fermi
pockets enclosing $\vec{k}_T$ in total implies a $\pi$ geometrical
(Berry's) phase [11]. In order to realize a $\pi$ Berry's phase the
surface bands must be spin-polarized and exhibit a partner switching
[11] dispersion behavior between a pair of $\vec{k}_T$. This means
that any pair of spin-polarized surface bands that are degenerate at
$\bar{\Gamma}$ must not re-connect at \={M}, or must separately
connect to the bulk valence and conduction band in between
$\bar{\Gamma}$ and \={M}. The partner switching behavior is realized
in Fig. 1(C) because the spin down band connects to and is
degenerate with different spin up bands at $\bar{\Gamma}$ and \={M}.
The partner switching behavior is realized in Fig. 2(A) because the
spin up and spin down bands emerging from $\bar{\Gamma}$ separately
merge into the bulk valence and conduction bands respectively
between $\bar{\Gamma}$ and \={M}.

We first investigate the spin properties of the topological
insulator phase. Spin-integrated ARPES [19] intensity maps of the
(111) surface states of insulating Bi$_{1-x}$Sb$_x$ taken at the
Fermi level ($E_F$) [Figs 1(D)\&(E)] show that a hexagonal FS
encloses $\bar{\Gamma}$, while dumbbell shaped FS pockets that are
much weaker in intensity enclose \={M}. By examining the surface
band dispersion below the Fermi level [Fig.1(F)] it is clear that
the central hexagonal FS is formed by a single band (Fermi crossing
1) whereas the dumbbell shaped FSs are formed by the merger of two
bands (Fermi crossings 4 and 5) [10].

This band dispersion resembles the partner switching dispersion
behavior characteristic of topological insulators. To check this
scenario and determine the topological index $\nu_0$, we have
carried out spin-resolved photoemission spectroscopy. Fig.1(G) shows
a spin-resolved momentum distribution curve taken along the
$\bar{\Gamma}$-\={M} direction at a binding energy $E_B$ = $-$25 meV
[Fig.1(G)]. The data reveal a clear difference between the spin-up
and spin-down intensities of bands 1, 2 and 3, and show that bands 1
and 2 have opposite spin whereas bands 2 and 3 have the same spin
(detailed analysis discussed later in text). The former observation
confirms that bands 1 and 2 form a spin-orbit split pair, and the
latter observation suggests that bands 2 and 3 (as opposed to bands
1 and 3) are connected above the Fermi level and form one band. This
is further confirmed by directly imaging the bands through raising
the chemical potential via doping [see supporting online material
(APPENDIX B) [22]]. Irrelevance of bands 2 and 3 to the topology is
consistent with the fact that the Fermi surface pocket they form
does not enclose any $\vec{k}_T$. Because of a dramatic intrinsic
weakening of signal intensity near crossings 4 and 5, and the small
energy and momentum splittings of bands 4 and 5 lying at the
resolution limit of modern spin-resolved ARPES spectrometers, no
conclusive spin information about these two bands can be drawn from
the methods employed in obtaining the data sets in Figs 1(G)\&(H).
However, whether bands 4 and 5 are both singly or doubly degenerate
does not change the fact that an odd number of spin-polarized FSs
enclose the $\vec{k}_T$, which provides evidence that
Bi$_{1-x}$Sb$_x$ has $\nu_0$ = 1 and that its surface supports a
non-trivial Berry's phase.

We investigated the quantum origin of topological order in this
class of materials. It has been theoretically speculated that the
novel topological order originates from the parities of the
electrons in pure Sb and not Bi [11,23]. It was also noted [20] that
the origin of the topological effects can only be tested by
measuring the spin-texture of the Sb surface, which has not been
measured. Based on quantum oscillation and magneto-optical studies,
the bulk band structure of Sb is known to evolve from that of
insulating Bi$_{1-x}$Sb$_x$ through the hole-like band at H rising
above $E_F$ and the electron-like band at $L$ sinking below $E_F$
[23]. The relative energy ordering of the $L_a$ and $L_s$ states in
Sb again determines whether the surface state pair emerging from
$\bar{\Gamma}$ switches partners [Fig.2(A)] or not [Fig.2(B)]
between $\bar{\Gamma}$ and \={M}, and in turn determines whether
they support a non-zero Berry's phase.

\begin{figure*}
\includegraphics[scale=0.62,clip=true, viewport=0.0in 0in 11.0in 6.0in]{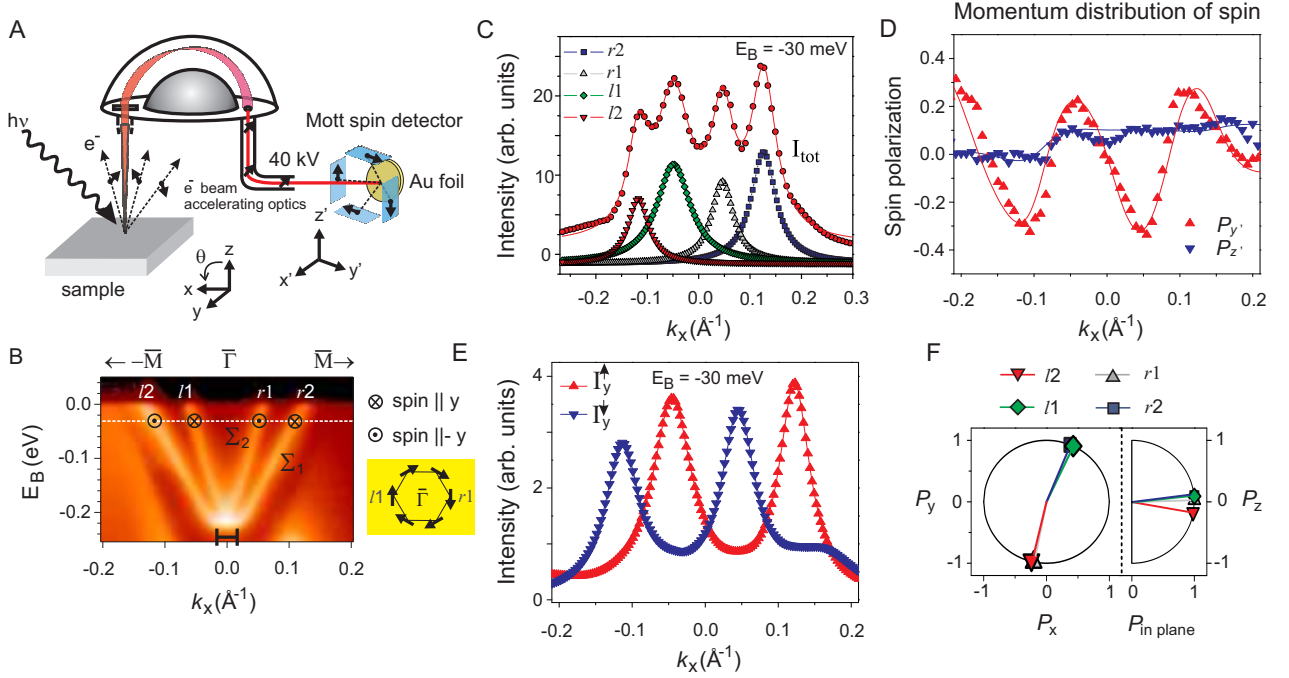}
\caption{Topological Spin-Texture and $\pi$ Berry's Phase in pure Sb (Antimony)}
\end{figure*}

In a conventional spin-orbit metal such as gold, a free-electron
like surface state is split into two parabolic spin-polarized
sub-bands that are shifted in $\vec{k}$-space relative to each other
[18]. Two concentric spin-polarized Fermi surfaces are created, one
having an opposite sense of in-plane spin rotation from the other,
that enclose $\bar{\Gamma}$. Such a Fermi surface arrangement, like
the schematic shown in figure 2(B), does not support a non-zero
Berry's phase because the $\vec{k}_T$ are enclosed an even number of
times (2 for most known materials).

However, for Sb, this is not the case. Figure 2(C) shows a
spin-integrated ARPES intensity map of Sb(111) from $\bar{\Gamma}$
to \={M}. By performing a systematic incident photon energy
dependence study of such spectra, previously unavailable with He
lamp sources [24], it is possible to identify two V-shaped surface
states (SS) centered at $\bar{\Gamma}$, a bulk state located near
$k_x$ = $-$0.25 \AA$^{-1}$ and resonance states centered about $k_x$
= 0.25 \AA$^{-1}$ and \={M} that are hybrid states formed by surface
and bulk states [19] (APPENDIX C [22]). An examination of the ARPES
intensity map of the Sb(111) surface and resonance states at $E_F$
[Fig.2(E)] reveals that the central surface FS enclosing
$\bar{\Gamma}$ is formed by the inner V-shaped SS only. The outer
V-shaped SS on the other hand forms part of a tear-drop shaped FS
that does \textit{not} enclose $\bar{\Gamma}$, unlike the case in
gold. This tear-drop shaped FS is formed partly by the outer
V-shaped SS and partly by the hole-like resonance state. The
electron-like resonance state FS enclosing \={M} does not affect the
determination of $\nu_0$ because it must be doubly spin degenerate
(APPENDIX D [22]). Such a FS geometry [Fig.2(G)] suggests that the
V-shaped SS pair may undergo a partner switching behavior expected
in Fig.2(A). This behavior is most clearly seen in a cut taken along
the $\bar{\Gamma}$-\={K} direction since the top of the bulk valence
band is well below $E_F$ [Fig.2(F)] showing only the inner V-shaped
SS crossing $E_F$ while the outer V-shaped SS bends back towards the
bulk valence band near $k_x$ = 0.1 \AA$^{-1}$ before reaching $E_F$.
The additional support for this band dispersion behavior comes from
tight binding surface calculations on Sb [Fig.2(D)], which closely
match with experimental data below $E_F$. Our observation of a
single surface band forming a FS enclosing $\bar{\Gamma}$ suggests
that pure Sb is likely described by $\nu_0$ = 1, and that its
surface may support a Berry's phase.

\begin{figure}[h]
\renewcommand{\thefigure}{S2}
\includegraphics[scale=0.35,clip=true, viewport=0.0in 0.85in 8.8in 8.8in]{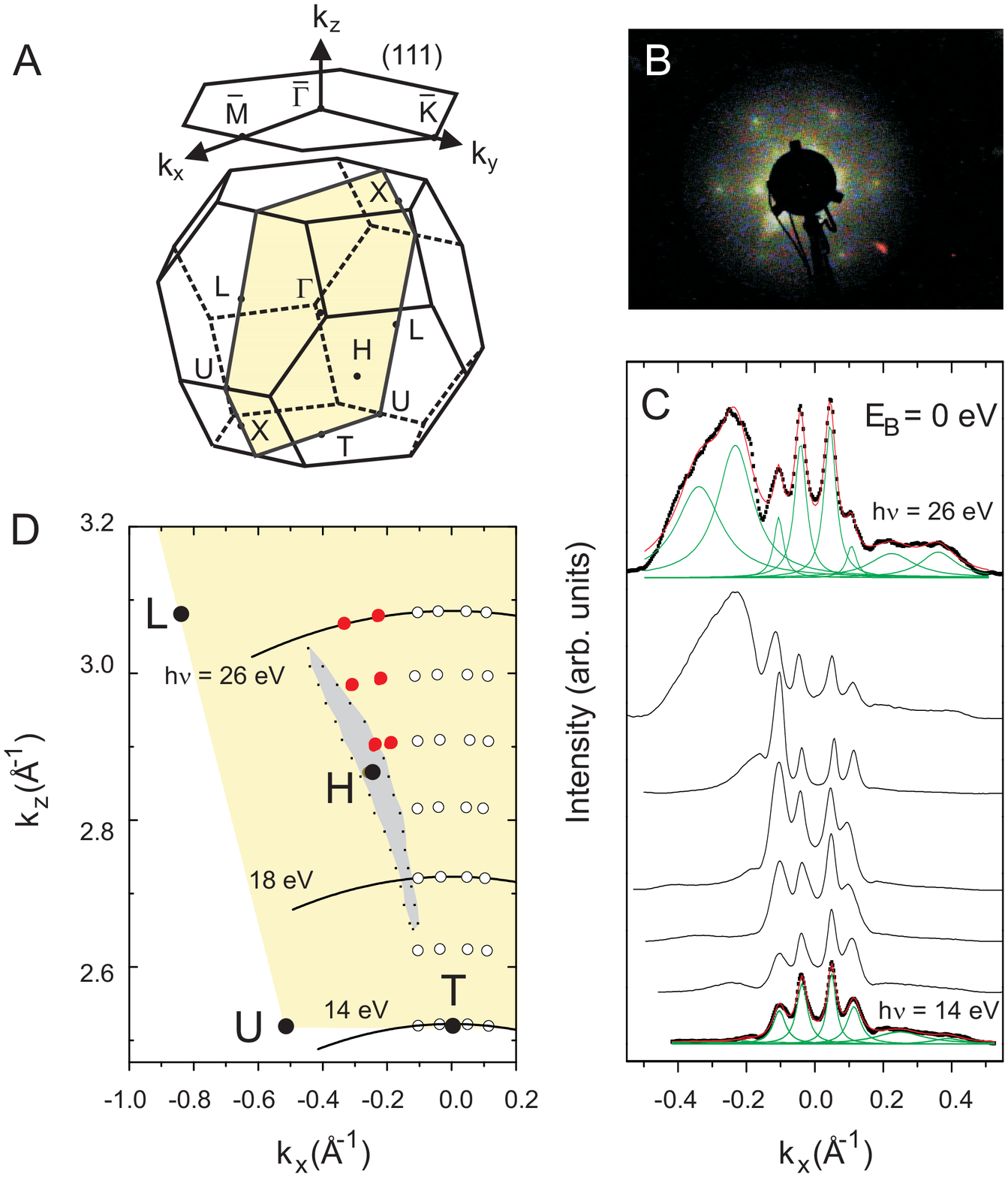}
\caption{\label{fig:Sb_FigS3} Incident energy and Brillouin zone space in pure Sb crystal.}
\end{figure}

Confirmation of a surface $\pi$ Berry's phase rests critically on a
measurement of the relative spin orientations (up or down) of the SS
bands near $\bar{\Gamma}$ so that the partner switching is indeed
realized, which cannot be done without spin resolution. Spin
resolution was achieved using a Mott polarimeter that measures two
orthogonal spin components of a photoemitted electron [27,28]. These
two components are along the $y'$ and $z'$ directions of the Mott
coordinate frame, which lie predominantly in and out of the sample
(111) plane respectively. Each of these two directions represents a
normal to a scattering plane defined by the photoelectron incidence
direction on a gold foil and two electron detectors mounted on
either side (left and right) [Fig.3(A)]. Strong spin-orbit coupling
of atomic gold is known to create an asymmetry in the scattering of
a photoelectron off the gold foil that depends on its spin component
normal to the scattering plane [28]. This leads to an asymmetry
between the left intensity ($I^L_{y',z'}$) and right intensity
($I^R_{y',z'}$) given by $A_{y',z'} =
(I^L_{y',z'}-I^R_{y',z'})/(I^L_{y',z'}+I^R_{y',z'})$, which is
related to the spin polarization $P_{y',z'} = (1/S_{eff})\times
A_{y',z'}$ through the Sherman function $S_{eff}$ = 0.085 [27,28].
Spin-resolved momentum distribution curve data sets of the SS bands
along the $-$\={M}-$\bar{\Gamma}$-\={M} cut at $E_B$ = $-$30 meV
[Fig.3(B)] are shown for maximal intensity. Figure 3(D) displays
both $y'$ and $z'$ polarization components along this cut, showing
clear evidence that the bands are spin polarized, with spins
pointing largely in the (111) plane. In order to estimate the full
3D spin polarization vectors from a two component measurement (which
is not required to prove the partner switching or the Berry's
phase), we fit a model polarization curve to our data following the
recent demonstration in Ref-[26], which takes the polarization
directions associated with each momentum distribution curve peak
[Fig.3(C)] as input parameters, with the constraint that each
polarization vector has length one (in angular momentum units of
$\hbar$/2). Our fitted polarization vectors are displayed in the
sample ($x,y,z$) coordinate frame [Fig.3(F)], from which we derive
the spin-resolved momentum distribution curves for the spin
components parallel ($I_y^{\uparrow}$) and anti-parallel
($I_y^{\downarrow}$) to the $y$ direction (APPENDIX B [22]) as shown
in figure 3(E). There is a clear difference in $I_y^{\uparrow}$ and
$I_y^{\downarrow}$ at each of the four momentum distribution curve
peaks indicating that the surface state bands are spin polarized
[Fig.3(E)], which is possible to conclude even without a full 3D
fitting. Each of the pairs $l2/l1$ and $r1/r2$ have opposite spin,
consistent with the behavior of a spin split pair, and the spin
polarization of these bands are reversed on either side of
$\bar{\Gamma}$ in accordance with the system being time reversal
symmetric $[E(\vec{k},\uparrow) = E(-\vec{k},\downarrow)]$
[Fig.3(F)]. The measured spin texture of the Sb(111) surface states
(Fig.3), together with the connectivity of the surface bands
(Fig.2), uniquely determines its belonging to the $\nu_0$ = 1 class.
Therefore the surface of Sb carries a non-zero ($\pi$) Berry's phase
via the inner V-shaped band and pure Sb can be regarded as the
parent metal of the Bi$_{1-x}$Sb$_x$ topological insulator class, in
other words, the topological order originates from the Sb wave
functions.

\begin{figure}[h]
\renewcommand{\thefigure}{S4}
\includegraphics[scale=0.27,clip=true, viewport=0.0in 0.45in 10.8in 7.5in]{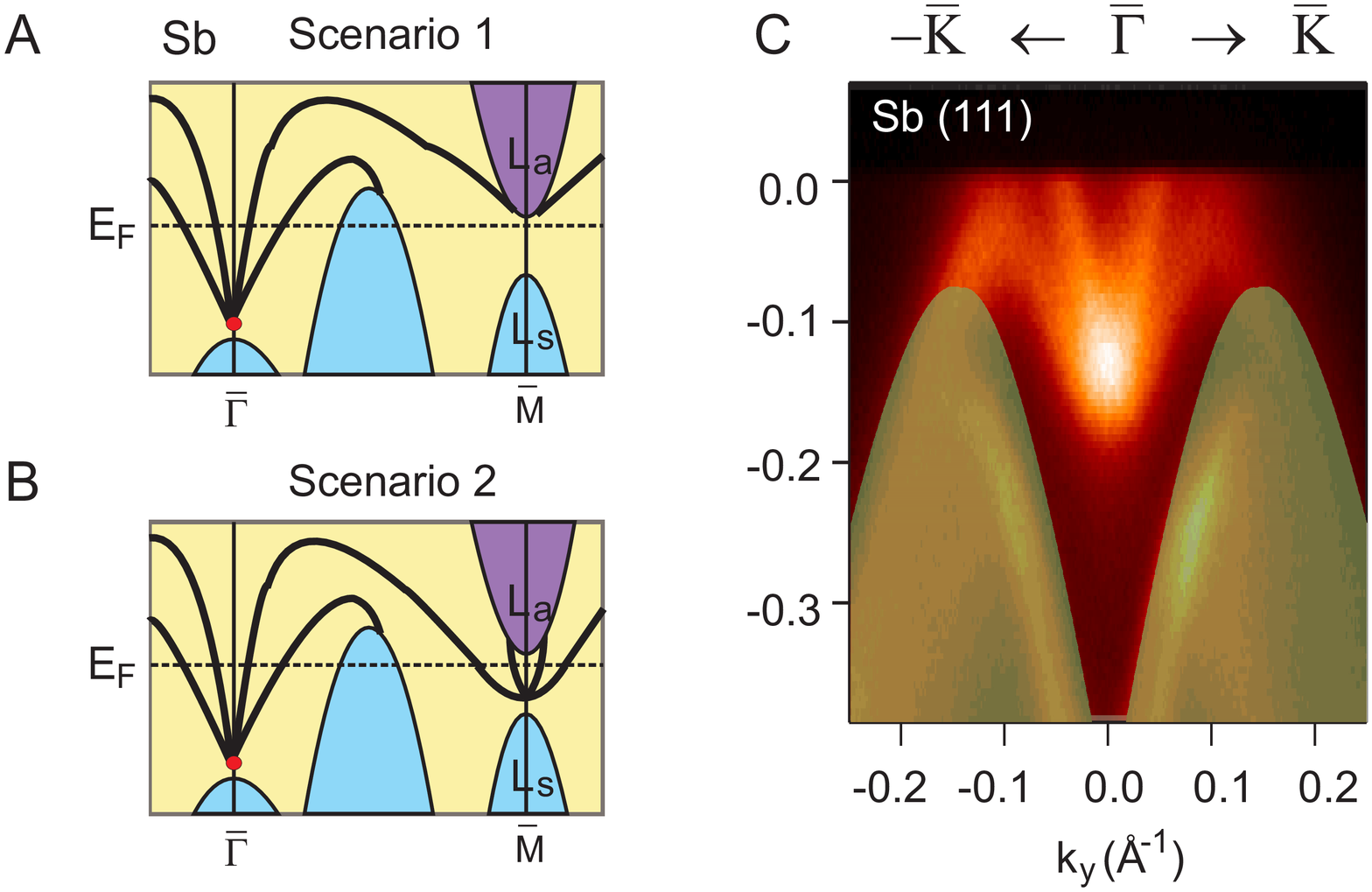}
\caption{\label{fig:Sb_FigS1} Surface band topologies in pure Sb.}
\end{figure}

\begin{figure}[h]
\renewcommand{\thefigure}{S6}
\includegraphics[scale=0.33,clip=true, viewport=0.0in 0in 11.0in 6.8in]{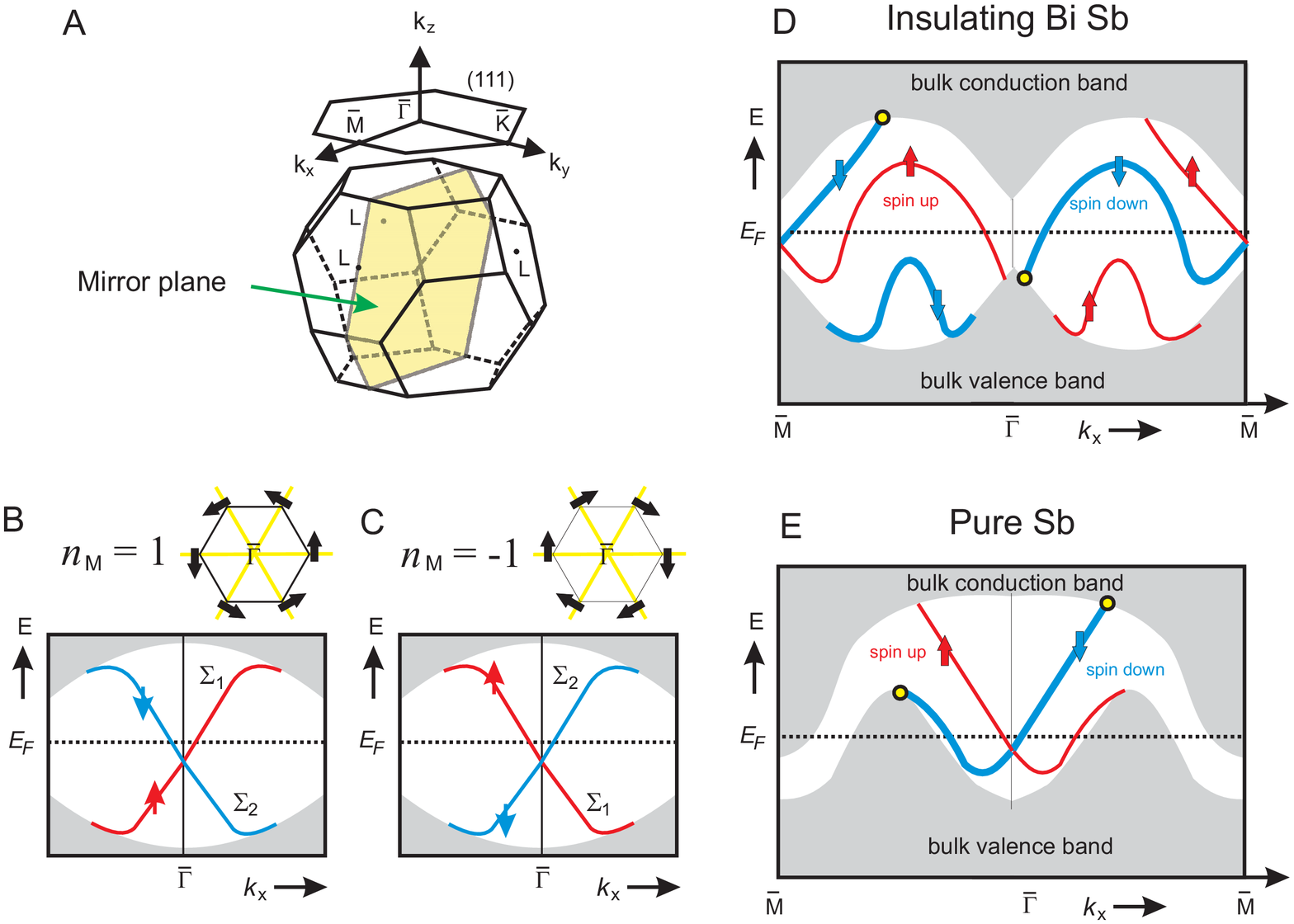}
\caption{\label{fig:Sb_FigS6} Z$_2$ Topology and Berry's Phase associated with the spin-polarized surface states in Sb and BiSb. The value of topological Mirror Chern number in Sb and BiSb suggest suggests left-handed chirality quantum number n$_M$ = -1.}
\end{figure}

Our spin polarized measurement methods (Fig.1 and 3) uncover a new
type of topological quantum number $n_M$ which provides information
about the chirality properties. Topological band theory suggests
that the bulk electronic states in the mirror ($k_y$ = 0) plane can
be classified in terms of a number $n_M$ (=$\pm$1) that describes
the handedness (either left or right handed) or chirality of the
surface spins which can be directly measured or seen in
spin-resolved experiments [20]. We now determine the value of $n_M$
from our data. From figure 1, it is seen that a single (one) surface
band, which switches partners at \={M}, connects the bulk valence
and conduction bands, so $|n_M|$ = 1 (APPENDIX F [22]). The sign of
$n_M$ is related to the direction of the spin polarization $\langle
\vec{P} \rangle$ of this band [20], which is constrained by mirror
symmetry to point along $\pm\hat{y}$. Since the central
electron-like FS enclosing $\bar{\Gamma}$ intersects six mirror
invariant points [see Fig.3(B)], the sign of $n_M$ distinguishes two
distinct types of handedness for this spin polarized FS. Figures
1(F) and 3 show that for both Bi$_{1-x}$Sb$_x$ and Sb, the surface
band that forms this electron pocket has $\langle \vec{P} \rangle
\propto -\hat{y}$ along the $k_x$ direction, suggesting a
left-handed rotation sense for the spins around this central FS thus
$n_M$ = $-$1. Therefore, both insulating Bi$_{1-x}$Sb$_x$ and pure
Sb possess equivalent chirality properties $-$ a definite spin
rotation sense (left-handed chirality, see Fig.3(B)) and a
topological Berry's phase.

These spin-resolved experimental measurements reveal an intimate and
straightforward connection between the topological numbers ($\nu_0$,
$n_M$) and the physical observables. The $\nu_0$ determines whether
the surface electrons support a non-trivial Berry's phase, and if
they do, the $n_M$ determines the spin handedness of the Fermi
surface that manifests this Berry's phase. The 2D Berry's phase is a
critical signature of topological order and is not realizable in
isolated 2D electron systems, nor on the surfaces of conventional
spin-orbit or exchange coupled magnetic materials. A non-zero
Berry's phase is known, theoretically, to protect an electron system
against the almost universal weak-localization behavior in their low
temperature transport [11,13] and is expected to form the key
element for fault-tolerant computation schemes [13,29], because the
Berry's phase is a geometrical agent or mechanism for protection
against quantum decoherence [30]. Its remarkable realization on the
Bi$_{1-x}$Sb$_x$ surface represents an unprecedented example of a 2D
$\pi$ Berry's phase, and opens the possibility for building
realistic prototype systems to test quantum computing modules. In
general, our results demonstrate that spin-ARPES is a powerful probe
of topological order and quantum spin Hall physics, which opens up a
new search front for topological materials for novel spin-devices
and fault-tolerant quantum computing.

\vspace{2cm}

\begin{center}
\textbf{Details of Materials and Methods}\\
\end{center}

Spin-integrated angle-resolved photoemission spectroscopy (ARPES)
measurements were performed with 14 to 30 eV photons on beam line
5-4 at the Stanford Synchrotron Radiation Laboratory, and with 28 to
32 eV photons on beam line 12 at the Advanced Light Source, both
endstations being equipped with a Scienta hemispherical electron
analyzer (see VG Scienta manufacturer website for instrument
specifications). Spin-resolved ARPES measurements were performed at
the SIS beam line at the Swiss Light Source using the COPHEE
spectrometer (31, p.15) with a single 40 kV classical Mott detector
and photon energies of 20 and 22 eV. The typical energy and momentum
resolution was 15 meV and 1.5\% of the surface Brillouin zone (BZ)
respectively at beam line 5-4, 9 meV and 1\% of the surface BZ
respectively at beam line 12, and 80 meV and 3\% of the surface BZ
respectively at SIS using a pass energy of 3 eV. The undoped and Te
doped Bi$_{1-x}$Sb$_{x}$ single crystal samples were each cleaved
from a boule grown from a stoichiometric mixture of high purity
elements. The boule was cooled from 650 to 270 $^{\circ}$C over a
period of 5 days and was annealed for 7 days at 270 $^{\circ}$C. Our
ARPES results were reproducible over many different sample batches.
Determination of the Sb compositions in Bi$_{1-x}$Sb$_x$ to 1\%
precision was achieved by bulk resistivity measurements, which are
very sensitive to Sb concentration (23), as well as scanning
electron microscopy analysis on a cleaved surface showing lateral
compositional homogeneity over the length scale of our ARPES photon
beam size. X-ray diffraction (XRD) measurements were used to check
that the samples were single phase, and confirmed that the single
crystals presented in this paper have rhombohedral A7 crystal
structure (point group R\={3}m). The XRD patterns of the cleaved
crystals exhibit only the (333), (666), and (999) peaks showing that
the naturally cleaved surface is oriented along the trigonal (111)
axis. Room temperature data were recorded on a Bruker D8
diffractometer using Cu K$\alpha$ radiation ($\lambda$=1.54 \AA) and
a diffracted beam monochromator. The in-plane crystal orientation
was determined by Laue x-ray diffraction prior to insertion into an
ultra high vacuum environment. Cleaving these samples \textit{in
situ} between 10 K and 55 K at chamber pressures less than 5
$\times10^{-11}$ torr resulted in shiny flat surfaces, characterized
\textit{in situ} by low energy electron diffraction (LEED) to be
clean and well ordered with the same symmetry as the bulk
[Fig.~\ref{fig:Sb_FigS3}(B)]. This is consistent with photoelectron
diffraction measurements that
show no substantial structural relaxation of the Sb(111) surface (32).\\

\begin{center}
\textbf{Methods of using incident photon energy modulated
ARPES to separate the bulk from surface electronic states of pure antimony (Sb)}\\
\end{center}

In this section we detail incident photon energy modulated ARPES
experiments on the low lying electronic states of single crystal
Sb(111), which we employ to isolate the surface from bulk-like
electronic bands over the entire BZ. Figure~\ref{fig:Sb_FigS3}(C)
shows momentum distributions curves (MDCs) of electrons emitted at
$E_F$ as a function of $k_x$ ($\parallel$ $\bar{\Gamma}$-\={M}) for
Sb(111). The out-of-plane component of the momentum $k_z$ was
calculated for different incident photon energies ($h\nu$) using the
free electron final state approximation with an experimentally
determined inner potential of 14.5 eV (37, 38). There are four peaks
in the MDCs centered about $\bar{\Gamma}$ that show no dispersion
along $k_z$ and have narrow widths of $\Delta k_x \approx$ 0.03
\AA$^{-1}$. These are attributed to surface states and are similar
to those that appear in Sb(111) thin films (37). As $h\nu$ is
increased beyond 20 eV, a broad peak appears at $k_x \approx$ -0.2
\AA$^{-1}$, outside the $k$ range of the surface states near
$\bar{\Gamma}$, and eventually splits into two peaks. Such a strong
$k_z$ dispersion, together with a broadened linewidth ($\Delta k_x
\approx$ 0.12 \AA$^{-1}$), is indicative of bulk band behavior, and
indeed these MDC peaks trace out a Fermi surface
[Fig.~\ref{fig:Sb_FigS3}(D)] that is similar in shape to the hole
pocket calculated for bulk Sb near H (36). Therefore by choosing an
appropriate photon energy (e.g. $\leq$ 20 eV), the ARPES spectrum at
$E_F$ along $\bar{\Gamma}$-\={M} will have contributions from only
the surface states. The small bulk electron pocket centered at L is
not
accessed using the photon energy range we employed [Fig.~\ref{fig:Sb_FigS3}(D)].\\

Now we describe the experimental procedure used to distinguish pure
surface states from resonant states on Sb(111) through their
spectral signatures. ARPES spectra along $\bar{\Gamma}$-\={M} taken
at three different photon energies are shown in
Fig.~\ref{fig:Sb_FigS4}. Near $\bar{\Gamma}$ there are two rather
linearly dispersive electron like bands that meet exactly at
$\bar{\Gamma}$ at a binding energy $E_B \sim$ -0.2 eV. This behavior
is consistent with a pair of spin-split surface bands that become
degenerate at the time reversal invariant momentum ($\vec{k}_T$)
$\bar{\Gamma}$ due to Kramers degeneracy. The surface origin of this
pair of bands is established by their lack of dependence on $h\nu$
[Fig.~\ref{fig:Sb_FigS4}(A)-(C)]. A strongly photon energy
dispersive hole like band is clearly seen on the negative $k_x$ side
of the surface Kramers pair, which crosses $E_F$ for $h\nu=$ 24 eV
and gives rise to the bulk hole Fermi surface near H
[Fig.~\ref{fig:Sb_FigS3}(D)]. For $h\nu\leq$ 20 eV, this band shows
clear back folding near $E_B \approx$ -0.2 eV indicating that it has
completely sunk below $E_F$. Further evidence for its bulk origin
comes from its close match to band calculations
[Fig.~\ref{fig:Sb_FigS3}(D)]. Interestingly, at photon energies such
as 18 eV where the bulk bands are far below $E_F$, there remains a
uniform envelope of weak spectral intensity near $E_F$ in the shape
of the bulk hole pocket seen with $h\nu$ = 24 eV photons, which is
symmetric about $\bar{\Gamma}$. This envelope does not change shape
with $h\nu$ suggesting that it is of surface origin. Due to its weak
intensity relative to states at higher binding energy, these
features cannot be easily seen in the energy distribution curves
(EDCs) in Fig.~\ref{fig:Sb_FigS4}(A)-(C), but can be clearly
observed in the MDCs shown in Fig.~\ref{fig:Sb_FigS3}(C) especially
on the positive $k_x$ side. Centered about the \={M} point, we also
observe a crescent shaped envelope of weak intensity that does not
disperse with $k_z$ [Fig.~\ref{fig:Sb_FigS4}(D)-(F)], pointing to
its surface origin. Unlike the sharp surface states near
$\bar{\Gamma}$, the peaks in the EDCs of the feature near \={M} are
much broader ($\Delta E \sim$80 meV) than the spectrometer
resolution (15 meV). The origin of this diffuse ARPES signal is not
due to surface structural disorder because if that were the case,
electrons at $\bar{\Gamma}$ should be even more severely scattered
from defects than those at \={M}. In fact, the occurrence of both
sharp and diffuse surface states originates from a $k$ dependent
coupling to the bulk. As seen in Fig.2(D) of the main text, the
spin-split Kramers pair near $\bar{\Gamma}$ lie completely within
the gap of the projected bulk bands near $E_F$ attesting to their
purely surface character. In contrast, the weak diffuse hole like
band centered near $k_x$ = 0.3 \AA$^{-1}$ and electron like band
centered near $k_x$ = 0.8 \AA$^{-1}$ lie completely within the
projected bulk valence and conduction bands respectively, and thus
their ARPES spectra exhibit the expected lifetime broadening due to
coupling with the underlying
bulk continuum (39).\\

\begin{center}
\textbf{Method of counting spin Fermi surface $\vec{k}_T$ enclosures in pure Sb}\\
\end{center}

\vspace{1cm}

In this section we give a detailed explanation of why the surface
Fermi contours of Sb(111) that overlap with the projected bulk Fermi
surfaces can be neglected when determining the $\nu_0$ class of the
material. Although the Fermi surface formed by the surface resonance
near \={M} encloses the $\vec{k}_T$ \={M}, we will show that this
Fermi surface will only contribute an even number of enclosures and
thus not alter the overall evenness or oddness of $\vec{k}_T$
enclosures. Consider some time reversal symmetric perturbation that
lifts the bulk conduction L$_a$ band completely above $E_F$ so that
there is a direct excitation gap at L. Since this perturbation
preserves the energy ordering of the L$_a$ and L$_s$ states, it does
not change the $\nu_0$ class. At the same time, the weakly surface
bound electrons at \={M} can evolve in one of two ways. In one case,
this surface band can also be pushed up in energy by the
perturbation such that it remains completely inside the projected
bulk conduction band [Fig.~\ref{fig:Sb_FigS1}(A)]. In this case
there is no more density of states at $E_F$ around \={M}.
Alternatively the surface band can remain below $E_F$ so as to form
a pure surface state residing in the projected bulk gap. However by
Kramers theorem, this SS must be doubly spin degenerate at \={M} and
its FS must therefore enclose \={M} twice
[Fig.~\ref{fig:Sb_FigS1}(B)]. In determining $\nu_0$ for
semi-metallic Sb(111), one can therefore neglect all segments of the
FS that lie within the projected areas of the bulk FS [Fig.2(G) of
main text] because they can only contribute an even number of FS
enclosures, which does not change the modulo 2 sum of $\vec{k}_T$
enclosures.

In order to further experimentally confirm the topologically
non-trivial surface band dispersion shown in figures 2(C) and (D) of
the main text, we show ARPES intensity maps of Sb(111) along the
-\={K}$-\bar{\Gamma}-$\={K} direction. Figure~\ref{fig:Sb_FigS1}(C)
shows that the inner V-shaped band that was observed along the
-\={M}$-\bar{\Gamma}-$\={M} direction retains its V-shape along the
-\={K}$-\bar{\Gamma}-$\={K} direction and continues to cross the
Fermi level, which is expected since it forms the central hexagonal
Fermi surface. On the other hand, the outer V-shaped band that was
observed along the -\={M}$-\bar{\Gamma}-$\={M} direction no longer
crosses the Fermi level along the -\={K}$-\bar{\Gamma}-$\={K}
direction, instead folding back below the Fermi level around $k_y$ =
0.1 \AA$^{-1}$ and merging with the bulk valence band
[Fig.~\ref{fig:Sb_FigS1}(C)]. This confirms that it is the
$\Sigma_{1(2)}$ band starting from $\bar{\Gamma}$ that connects to
the bulk valence (conduction) band, in agreement with the
calculations shown in figure 2(D) of the main text.
\\
\\

\begin{center}
\textbf{Investigation of the robustness of Sb spin states under random field perturbations introduced by Bi substitutional disorder}\\
\end{center}

The predicted topological protection of the surface states of Sb
implies that their metallicity cannot be destroyed by weak time
reversal symmetric perturbations. In order to test the robustness of
the measured gapless surface states of Sb, we introduce such a
perturbation by randomly substituting Bi into the Sb crystal matrix
(APPENDIX A). Another motivation for performing such an experiment
is that the formalism developed by Fu and Kane (41) to calculate the
$Z_2$ topological invariants relies on inversion symmetry being
present in the bulk crystal, which they assumed to hold true even in
the random alloy Bi$_{1-x}$Sb$_x$. However, this formalism is simply
a device for simplifying the calculation and the non-trivial
$\nu_0=1$ topological class of Bi$_{1-x}$Sb$_x$ is predicted to hold
true even in the absence of inversion symmetry in the bulk crystal
(41). Therefore introducing light Bi substitutional disorder into
the Sb matrix is also a method to examine the effects of alloying
disorder and possible breakdown of bulk inversion symmetry on the
surface states of Sb(111). We have performed spin-integrated ARPES
measurements on single crystals of the random alloy
Sb$_{0.9}$Bi$_{0.1}$. Figure~\ref{fig:Sb_FigS5} shows that both the
surface band dispersion along $\bar{\Gamma}$-\={M} as well as the
surface state Fermi surface retain the same form as that observed in
Sb(111), and therefore the `topological metal' surface state of
Sb(111) fully survives the alloy disorder. Since Bi alloying is seen
to only affect the band structure of Sb weakly, it is reasonable to
assume that the topological order is preserved between Sb and
Bi$_{0.91}$Sb$_{0.09}$
as we observed.\\

\vspace{1cm}

\vspace{1cm}


\newpage

\textbf{FIG. 1.  Theoretical spin spectrum of a topological
insulator and spin-resolved spectroscopy results.} (A) Schematic
sketches of the bulk Brillouin zone (BZ) and (111) surface BZ of the
Bi$_{1-x}$Sb$_x$ crystal series. The high symmetry points
(L,H,T,$\Gamma$,$\bar{\Gamma}$,\={M},\={K}) are identified. (B)
Schematic of Fermi surface pockets formed by the surface states (SS)
of a topological insulator that carries a Berry's phase. (C) Partner
switching band structure topology: Schematic of spin-polarized SS
dispersion and connectivity between $\bar{\Gamma}$ and \={M}
required to realize the FS pockets shown in panel-(B). $L_a$ and
$L_s$ label bulk states at $L$ that are antisymmetric and symmetric
respectively under a parity transformation (see text). (D)
Spin-integrated ARPES intensity map of the SS of
Bi$_{0.91}$Sb$_{0.09}$ at $E_F$. Arrows point in the measured
direction of the spin. (E) High resolution ARPES intensity map of
the SS at $E_F$ that enclose the \={M}$_1$ and \={M}$_2$ points.
Corresponding band dispersion (second derivative images) are shown
below. The left right asymmetry of the band dispersions are due to
the slight offset of the alignment from the
$\bar{\Gamma}$-\={M}$_1$(\={M}$_2$) direction. (F) Surface band
dispersion image along the $\bar{\Gamma}$-\={M} direction showing
five Fermi level crossings. The intensity of bands 4,5 is scaled up
for clarity (the dashed white lines are guides to the eye). The
schematic projection of the bulk valence and conduction bands are
shown in shaded blue and purple areas. (G) Spin-resolved momentum
distribution curves presented at $E_B$ = $-$25 meV showing single
spin degeneracy of bands at 1, 2 and 3. Spin up and down correspond
to spin pointing along the +$\hat{y}$ and -$\hat{y}$ direction
respectively. (H) Schematic of the spin-polarized surface FS
observed in our experiments. It is consistent with a $\nu_0$ = 1
topology (compare (B) and (H)).

\vspace{1cm}

\textbf{FIG. 2.  Topological character of pure Sb revealed on the
(111) surface states.} Schematic of the bulk band structure (shaded
areas) and surface band structure (red and blue lines) of Sb near
$E_F$ for a (A) topologically non-trivial and (B) topological
trivial (gold-like) case, together with their corresponding surface
Fermi surfaces are shown. (C) Spin-integrated ARPES spectrum of
Sb(111) along the $\bar{\Gamma}$-\={M} direction. The surface states
are denoted by SS, bulk states by BS, and the hole-like resonance
states and electron-like resonance states by h RS and e$^-$ RS
respectively. (D) Calculated surface state band structure of Sb(111)
based on the methods in [20,25]. The continuum bulk energy bands are
represented with pink shaded regions, and the lines show the
discrete bands of a 100 layer slab. The red and blue single bands,
denoted $\Sigma_1$ and $\Sigma_2$, are the surface states bands with
spin polarization $\langle \vec{P} \rangle \propto +\hat{y}$ and
$\langle \vec{P} \rangle \propto -\hat{y}$ respectively. (E) ARPES
intensity map of Sb(111) at $E_F$ in the $k_x$-$k_y$ plane. The only
one FS encircling $\bar{\Gamma}$ seen in the data is formed by the
inner V-shaped SS band seen in panel-(C) and (F). The outer V-shaped
band bends back towards the bulk band best seen in data in
panel-(F). (F) ARPES spectrum of Sb(111) along the
$\bar{\Gamma}$-\={K} direction shows that the outer V-shaped SS band
merges with the bulk band. (G) Schematic of the surface FS of
Sb(111) showing the pockets formed by the surface states (unfilled)
and the resonant states (blue and purple). The purely surface state
Fermi pocket encloses only one Kramers degenerate point
($\vec{k}_T$), namely, $\bar{\Gamma}$(=$\vec{k}_T$), therefore
consistent with the $\nu_0$ = 1 topological classification of Sb
which is different from Au (compare (B) and (G)). As discussed in
the text, the hRS and e$^-$RS count trivially.

\vspace{1cm}

\textbf{FIG. 3.  Spin-texture of topological surface states and
chirality.} (A) Experimental geometry of the spin-resolved ARPES
study. At normal emission ($\theta$ = 0$^{\circ}$), the sensitive
$y'$-axis of the Mott detector is rotated by 45$^{\circ}$ from the
sample $\bar{\Gamma}$ to $-$\={M} ($\parallel -\hat{x}$) direction,
and the sensitive $z'$-axis of the Mott detector is parallel to the
sample normal ($\parallel \hat{z}$). (B) Spin-integrated ARPES
spectrum of Sb(111) along the $-$\={M}-$\bar{\Gamma}$-\={M}
direction. The momentum splitting between the band minima is
indicated by the black bar and is approximately 0.03 \AA$^{-1}$. A
schematic of the spin chirality of the central FS based on the
spin-resolved ARPES results is shown on the right. (C) Momentum
distribution curve of the spin averaged spectrum at $E_B$ = $-$30
meV (shown in (B) by white line), together with the Lorentzian peaks
of the fit. (D) Measured spin polarization curves (symbols) for the
detector $y'$ and $z'$ components together with the fitted lines
using the two-step fitting routine [26]. (E) Spin-resolved spectra
for the sample $y$ component based on the fitted spin polarization
curves shown in (D). Up (down) triangles represent a spin direction
along the +(-)$\hat{y}$ direction. (F) The in-plane and out-of-plane
spin polarization components in the sample coordinate frame obtained
from the spin polarization fit. Overall spin-resolved data and the
fact that the surface band that forms the central electron pocket
has $\langle \vec{P} \rangle \propto -\hat{y}$ along the +$k_x$
direction, as in (E), suggest a left-handed chirality (schematic in
(B) and see text for details).

\vspace{1cm}

\textbf{Fig. S2.} (\textbf{A}) Schematic of the bulk BZ of Sb and
its (111) surface BZ. The shaded region denotes the momentum plane
in which the following ARPES spectra were measured. (\textbf{B})
LEED image of the \textit{in situ} cleaved (111) surface exhibiting
a hexagonal symmetry. (\textbf{C}) Select MDCs at $E_F$ taken with
photon energies from 14 eV to 26 eV in steps of 2 eV, taken in the
$TXLU$ momentum plane. Peak positions in the MDCs were determined by
fitting to Lorentzians (green curves). (\textbf{D}) Experimental 3D
bulk Fermi surface near H (red circles) and 2D surface Fermi surface
near $\bar{\Gamma}$ (open circles) projected onto the $k_x$-$k_z$
plane, constructed from the peak positions found in (\textbf{C}).
The $k_z$ values are determined using calculated constant $h\nu$
contours (black curves) (see APPENDIX C text). The shaded gray
region is the theoretical hole Fermi surface calculated in (36).

\vspace{1cm}

\textbf{Fig. S4.} (\textbf{A}) Schematic of the surface band
structure of Sb(111) under a time reversal symmetric perturbation
that lifts the bulk conduction (L$_a$) band above the Fermi level
($E_F$). Here the surface bands near \={M} are also lifted completed
above $E_F$. (\textbf{B}) Alternatively the surface band near \={M}
can remain below $E_F$ in which case it must be doubly spin
degenerate at \={M}. (\textbf{C}) ARPES intensity plot of the
surface states along the -\={K}$-\bar{\Gamma}-$\={K} direction. The
shaded green regions denote the theoretical projection of the bulk
valence bands, calculated using the full potential linearized
augmented plane wave method using the local density approximation
including the spin-orbit interaction (method described in 40). Along
this direction, it is clear that the outer V-shaped surface band
that was observed along the -\={M}$-\bar{\Gamma}-$\={M} now merges
with the bulk valence band.

\vspace{1cm}

\textbf{Fig. S6.} Implications of k-space mirror symmetry on the
surface spin states. (\textbf{A}) 3D bulk Brillouin zone and the
mirror plane in reciprocal space. (\textbf{B}) Schematic spin
polarized surface state band structure for a mirror Chern number
($n_M$) of +1 and (\textbf{C}) -1. Spin up and down mean parallel
and anti-parallel to $\hat{y}$ respectively. The upper (lower)
shaded gray region corresponds to the projected bulk conduction
(valence) band. The hexagons are schematic spin polarized surface
Fermi surfaces for different $n_M$, with yellow lines denoting the
mirror planes. (\textbf{D}) Schematic representation of surface
state band structure of insulating Bi$_{1-x}$Sb$_x$ and (\textbf{E})
semi metallic Sb both showing a $n_M = -1$ topology. Yellow circles
indicate where the spin down band (bold) connects the bulk valence
and conduction bands.

\end{document}